\documentstyle[preprint,prd,aps,epsf]{revtex}
\tightenlines
\begin{document}
\draft
\title{Out-of-equilibrium Photon Production from  \\Disoriented Chiral Condensates
 }
\author{Da-Shin Lee\footnote{E-mail address: {\tt dslee@mail.ndhu.edu.tw}}}
\address{Department of Physics, National Dong Hwa University, Hua-Lien, 
Taiwan, R.O.C.}
\author{Kin-Wang Ng\footnote{E-mail address: {\tt nkw@phys.sinica.edu.tw}}}
\address{Institute of Physics, Academia Sinica, Taipei, Taiwan, R.O.C.}
\maketitle

\begin{abstract}
We study the production of photons  through the non-equilibrium relaxation 
of a disoriented chiral condensate.
We propose that to search for non-equilibrium
photons in the direct photon measurements of heavy-ion collisions can be a
potential test of the formation of disoriented chiral condensates.
\end{abstract}
\vspace{2pc}
\vspace{2pc}

Recently, there have been many  investigations into the formation of  `` Disoriented Chiral Condensates'' (DCCs) following relativistic heavy ion collisions proposed by Bjorken et al. for a navel signature for the chiral phase transition\cite{bjorken1,wilczek,dcc,cooper,boydcc}. The generally accepted picture of heavy  energetic nuclei collisions is based on the Bjorken's scenario. In the collisions of these highly Lorentz contracted nuclei, they essentially pass through each other, leaving behind a hot plasma in the central rapidity region with large energy  density corresponding to temperature above $200 ~{\rm MeV}$ where the chiral symmetry is restored. This plasma then cools down via rapid hydrodynamic expansion through the chiral phase transition during which the long-wavelength fluctuations become unstable and grow due to the spinodal instabilities \cite{wilczek,cooper,boydcc}. The growth of these unstable modes results in the formation of DCCs. The DCCs are the correlated regions of space-time where
 the chiral order parameter of QCD is chirally rotated  from its usual orientation in isospin space.
 Subsequent relaxation of such DCCs to the true QCD vacuum is expected to radiate copious soft pions with a navel distribution  in the ratio of neutral to charged pions, $f ~(P(f) \approx 1/ \sqrt{f} )$,  which could be a potential experimental signature of the chiral phase transition observable at RHIC and LHC.    However, since these emitted pions will undergo the strong final interaction, the signals for the distribution $P(f)$ may be severely masked and become indistinguishable from the background.
It then becomes important to study other possible signatures of DCCs that would be less affected by the final state interaction. 
Electromagnetic probe such as photon and lepton with longer mean free path in the medium  than the pions serves as a good candidate and can reveal more detailed non-equilibrium information on the DCCs with minimal distortion \cite{wang,boy1,boy2}.

                                                                                                                                                                                                                                                                                                                                                                                                                        Minakawa and Muller \cite{muller} have  recently suggested  that the presence of strong electromagnetic
fields in relativistic heavy ion collisions induces  a quasi-instantaneous ``kick'' to the field configuration along the $\pi^0$ direction such that it is plausible  that the chiral order parameter in the DCC domains, if formed, will acquire a component in the direction of the neutral pion.  
In this Letter, we  consider  the production of photons  through the non-equilibrium relaxation of a DCC within which  the chiral order parameter  initially has a non-vanishing expectation value along the $\pi^{0}$ direction and subsequently oscillates around the minimum of the effective potential. 
Our aim is to understand how the photons can be produced  from this oscillating  $\pi^{0}$ field via  the dynamics of parametric amplification as well as 
spinodal instabilities.  
In Ref.~\cite{boy2}, Boyanovsky et al. have extensively studied the  photon production
from the low energy coupling of the neutral pion to photon
via the ${\rm U}_{\rm A} (1)$ anomalous vertex.  
They have found that for large initial amplitudes of the $\pi^0$ field 
photon production is enhanced by parametric amplification.
These processes are non-perturbative with a large contribution during
the non-equilibrium stages of the evolution and result in a distinct
distribution of the produced photons.
Here we will take into account another dominant contribution that also involves  the dynamics of $\pi^{0}$ due to the decay of the vector meson through the 
electromagnetic vertex.  Although  the corresponding   dimensionless effective coupling  involving  the vector meson is quite small perturbatively, as we will see later, in fact, for the large amplitude oscillations of the $\pi^{0}$ mean field, the  contribution to the photon production  is of the same order of  magnitude as the anomalous interaction.

 The relevant  phenomenological   Lagrangian density is given by
\begin{equation}
{\cal L}= {\cal L}_\sigma+ {\cal L}_{\gamma} +{\cal L}_{\pi^0 \gamma \gamma} +{\cal L}_V + {\cal L}_{V \pi \gamma},
\label{lagrangian}
\end{equation}
where
\begin{eqnarray}
{\cal L}_\sigma &=& {1\over2} \partial^\mu {\vec\Phi}\cdot \partial_\mu
             {\vec\Phi} - {1\over2} m^2 (t)
             {\vec\Phi}\cdot {\vec\Phi}
             - \lambda \left({\vec\Phi}\cdot{\vec\Phi}\right)^2 + h\sigma, \\
{\cal L}_{\gamma}+ {\cal L}_{\pi^0 \gamma \gamma} &=& -{1\over4}  F^{\mu\nu} F_{\mu\nu} + \frac{e^2}{32\pi^2}\frac{\pi^0}{f_\pi}
             \epsilon^{\alpha\beta\mu\nu} F_{\alpha\beta} F_{\mu\nu}
, \\
{\cal L}_V + {\cal L}_{V \pi \gamma} &=& -{1\over4}  V^{\mu\nu} V_{\mu\nu} - \frac{1}{2}  m_V V^{\mu} V_{\mu} + \frac{e \lambda_{V}}{4 m _{\pi}}
             \epsilon^{\alpha\beta\mu\nu} F_{\alpha\beta} V_{\mu\nu} \pi^{0},
\end{eqnarray}
where $F_{\mu\nu}=\partial_\mu A_\nu - \partial_\nu A_\mu$ is the electromagnetic field-strength tensor, and $V_{\mu\nu}=\partial_\mu V_\nu - \partial_\nu V_\mu$ is the field-strength  tensor of the vector meson with mass $m_{V}$. In addition,  
${\vec\Phi}=(\sigma,\pi^0,{\vec\pi})$ is an $O(N+1)$ vector with 
$\vec\pi=(\pi^1,\pi^2,...,\pi^{N-1})$ representing the $N-1$ pions. 
This Lagrangian without the vector meson piece is the model considered in
Ref.~\cite{boy2}. In this work, we will follow closely the formalism developed
there. Likewise, we are going to  ignore the hydrodynamical expansion and adopt the
simple ``quench'' phase transition from an initial thermodynamic equilibrium state at a temperature ($T_i$) higher than the critical temperature ($T_c$) for the chiral phase transition cooled instantaneously to zero temperature.  This ``quench'' scenario, which has been widely used in the study of non-equilibrium phenomena of DCCs \cite{boydcc,wang,boy1,boy2}, can capture the qualitative features of this non-equilibrium problem  and allow a concrete analytical calculation. 
Thus, we take~\cite{boydcc,boy1,boy2}
\begin{equation}
m^2 (t)= \frac{m^2_{\sigma}}{2} \left[ \frac{T_i^2}{T_c^2} \Theta (-t) -1 \right], ~~ T_i > T_c.
\end{equation}
The parameters can be determined  by the low-energy pion physics as follows:
\begin{eqnarray}
&& m_{\sigma} \approx 600 ~{\rm MeV}, ~~ f_{\pi} \approx 93 ~{\rm MeV}, ~~ {\lambda} \approx 4.5, ~~ T_c \approx 200 ~{\rm MeV}, \nonumber \\
&& h \approx (120~ {\rm MeV})^3,~~ m_{V} \approx 782 ~{\rm MeV},~~ \lambda_{V} \approx 0.36,
\label{parameter}
\end{eqnarray}
where $V$ is identified as the $\omega$ meson, and 
the coupling $\lambda_{V}$ is  obtained  from the $ \omega \rightarrow \pi^0 \gamma$
decay width \cite{david}.

Since we are only interested in the photon production, we can integrate out the vector meson to obtain the effective Lagrangian density that contains the relevant degrees of freedom given by
\begin{equation}
{\cal L}_{\rm eff} 
=  {\cal L}_\sigma+ {\cal L}_{\gamma} +{\cal L}_{\pi^0 \gamma \gamma}
             -\frac{e^2\lambda_V^2}{8m_\pi^2 m_V^2}
  \epsilon^{\mu\nu\lambda\delta}\epsilon^{\alpha\beta\gamma}_{~~~\delta}~
              \partial_\lambda \pi^0 \partial_\gamma \pi^0
             F_{\mu\nu}F_{\alpha\beta},                
\end{equation}
where the higher derivative terms are dropped out. At this point it must be noticed that  here we have assumed the validity of the low-energy effective vertices. The effective vertices that account for the above mentioned processes may be modified in the strongly out of equilibrium situation \cite{boy4}.  In fact, one should obtain the non-equilibrium vertices by integrating out the quark fields and the vector meson in the context of the fully non-equilibrium formalism that we are currently studying in detail. 

Before proceeding further, let us discuss the  qualitative features of the relative importance  of  these two effective  couplings to the photon production .  For small amplitude oscillations of the $\pi^{0}$  mean field  (e.g. smaller than the mass of the ${\pi^{0}}$), from the naive perturbation argument with the parameters in Eq.~(\ref{parameter}),  we expect that the effect of photon production from the coupling of  $(\partial\pi^0 {\tilde F})^2 $   is one order of magnitude smaller than that of  the anomalous interaction. However,
for large amplitude oscillations, the non-linearity results in the $(\partial\pi^0 {\tilde F})^2 $  term being of the same order of magnitude as  the  anomalous interaction, and moreover this non-linear effect will make the photon production mechanism  due to these two couplings more effective as we can see later.

Following the non-equilibrium quantum field theory that requires a path integral representation along the complex contour in time \cite{boyneq}, the non-equilibrium Lagrangian density is given by
\begin{equation}
{\cal L}_{\rm neq} ={\cal L}_{\rm eff} [\Phi^+, A_{\mu}^+ ] -{\cal L}_{\rm eff} [\Phi^-, A_{\mu}^- ],
\end{equation}
where $ + (-)$ denotes the forward (backward) time branches. The non-equilibrium equations of motion are obtained via the tadpole method.    
As mentioned before, the situation of interest to us is a DCC in which both 
the $\sigma $ and $\pi^0$ fields acquire the vacuum expectation values. We then shift 
$\sigma$ and $\pi^0$ by their  expectation values described by the initial non-equilibrium states specified later,
\begin{eqnarray}
\sigma(\vec x,t) &=& \phi(t) + \chi(\vec x,t), ~~ \phi(t)=\langle \sigma(\vec x,t) \rangle, \nonumber \\
\pi^{0}(\vec x,t) &=& \zeta(t) + \psi(\vec x,t),~~~~ \zeta(t)=\langle \pi^0(\vec x,t) \rangle,
\end{eqnarray}
with the tadpole conditions,
\begin{equation}
\langle \chi(\vec x,t) \rangle = 0,~~\langle \psi(\vec x,t) \rangle = 0, ~~
\langle \vec{\pi}(\vec x,t) \rangle = 0.
\label{tad}
\end{equation}
This tadpole conditions will be imposed to all orders in the corresponding 
expansion.
 In order to derive the non-equilibrium evolution equations that 
incorporate  quantum fluctuation effects 
from the strong $\sigma-\pi$   interactions, we will  use the large-$N$ limit to provide a consistent, non-perturbative framework to study this dynamics.  
To leading order in the $1/N$ expansion, following the  
Hartree factorizations (Eqs.~(2.7)-(2.9) of Ref.~\cite{boy2})
implemented for both $\pm$ components,
the Lagrangian then becomes
\begin{eqnarray}
 &&{\cal L}_{\rm eff}\left[ \phi+\chi^+, \zeta + \psi^+, {\vec \pi}^+,  A_\mu^+\right] -
               {\cal L}_{\rm eff} \left[  \phi +\chi^-, \zeta + \psi^-, {\vec \pi}^-  
              ,A_\mu^-\right]  \nonumber \\
&& =\left\{ \frac{1}{2} (\partial\chi^+)^2 +\frac{1}{2} (\partial\psi^+)^2
      +\frac{1}{2} (\partial{\vec \pi}^+)^2
     - U_{1} (t) \chi^+ - U_{2} (t) \psi^+   
      \right. \nonumber\\
&& -\frac{1}{2} M_{\chi}^{2}(t) \chi^{+2}  -\frac{1}{2} M_{\psi}^{2}(t)    
      \psi^{+2}   -\frac{1}{2} M_{\vec{\pi}}^{2}(t) {\vec{\pi}}^{+2}
      -{1\over 4}F^{+}_{\mu\nu}F^{+\mu\nu} 
+\frac{e^2  }{32 \pi^2  f_{\pi}}\zeta(t)  \epsilon^{\alpha\beta\mu\nu} F^+_{\alpha\beta} F^+_{\mu\nu} \nonumber \\
&& +\frac{e^2  }{32 \pi^2  f_{\pi}}\psi^+  \epsilon^{\alpha\beta\mu\nu} F^+_{\alpha\beta} F^+_{\mu\nu}
  -\frac{e^2 \lambda^2_V }{8 m^2_V m^2_{\pi} } (\dot\zeta(t))^2 
  \epsilon^{\mu\nu 0 \delta}
 \epsilon^{\alpha\beta 0}_{~~~\delta}~
         F^+_{\mu\nu}F^+_{\alpha\beta}   + \frac{e^2 \lambda^2_V }{4 m^2_V m^2_{\pi} } \ddot\zeta(t) \psi^+ 
 \epsilon^{\mu\nu 0 \delta}
 \epsilon^{\alpha\beta 0}_{~~~\delta}~
         F^+_{\mu\nu}F^+_{\alpha\beta}  \nonumber \\
&&+ \frac{e^2 \lambda^2_V }{4 m^2_V m^2_{\pi} } \dot\zeta(t) \psi^+ 
 \epsilon^{\mu\nu 0 \delta}
 \epsilon^{\alpha\beta \sigma}_{~~~\delta}~
       \partial_{\sigma}  F^+_{\mu\nu}F^+_{\alpha\beta}  
\left. - \frac{e^2 \lambda^2_V }{8 m^2_V m^2_{\pi} }  \epsilon^{\mu\nu \lambda \delta}
 \epsilon^{\alpha\beta \gamma}_{~~~\delta}~
\partial_{\lambda} \psi^+  \partial_{\gamma} \psi^+ 
F^+_{\mu\nu}F^+_{\alpha\beta} 
  \right\} 
-\left\{ + \rightarrow -\right\}, \nonumber \\
\end{eqnarray} 
where 
\begin{eqnarray}
 U_1 (t) &=& \ddot\phi(t)+ \left[m^2 (t) + 4\lambda \phi^2 (t)+ 
             4\lambda \zeta^2 (t)+4 \lambda\Sigma(t)\right] \phi(t)-h ,  \nonumber \\
U_2 (t) &=& \ddot\zeta(t)+ \left[m^2 (t) + 4\lambda \phi^2 (t)+ 
             4\lambda \zeta^2 (t)+4 \lambda \Sigma(t)\right] \zeta(t),  \nonumber \\
M_{\chi}^{2}(t) &=&  m^2 (t) + 12 \lambda \phi^2 (t)+ 
                  4\lambda \zeta^2 (t)+4 \lambda\Sigma(t)
                      \nonumber \\
M_{\psi}^{2}(t) &=& m^2 (t) + 4 \lambda \phi^2 (t)+ 12 \lambda
              \zeta^2 (t)+4 \lambda\Sigma(t), 
                      \nonumber \\
M_{\vec{\pi}}^{2}(t) &=& m^2 (t) + 4 \lambda \phi^2 (t)+ 4 \lambda
              \zeta^2 (t)+4 \lambda\Sigma(t), 
                      \nonumber \\
\Sigma(t)&=&  \langle{\vec{\pi}}^2\rangle(t) -\langle{\vec{\pi}}\rangle(0).
\end{eqnarray}
The  expectation value described by the initial non-equilibrium states will be determined self-consistently.
Here, we have performed a subtraction of $\langle{\vec{\pi}}^2\rangle(t)$ at $t=0$ 
absorbing $\langle{\vec{\pi}}^2\rangle(0)$ into the finite renormalization of the mass term.  

Since we consider the direct 
photon production driven by   the time dependent oscillating   field in which the photon does not appear in the intermediate states. It proves to be convenient to choose the Coulomb gauge that  contains  physical
degrees of freedom
without any other redundant fields~\cite{boy2}. 
With the above Hartree-factorized Lagrangian in the Coulomb gauge, 
following the tadpole conditions, we can obtain the full one-loop equations of motion while we treat the weak electromagnetic coupling perturbatively:
\begin{eqnarray}
&&\ddot\phi(t) + \left[m^2 (t)+ 4 \lambda \phi^2 (t)+ 4 \lambda \zeta^2(t)+ 4 \lambda \Sigma(t)\right] \phi(t)- h=0, \nonumber \\
&&\ddot\zeta(t) + \left[m^2 (t)+ 4 \lambda \phi^2 (t)+4 \lambda\zeta^2(t)+ 4 \lambda\Sigma(t)\right] \zeta(t)- 
\frac{e^2}{32 \pi^2  f_{\pi}}\epsilon^{\alpha\beta\mu\nu} \langle F_{\alpha\beta}   F_{\mu\nu} \rangle(t) 
 \nonumber \\
&&~~~~~~~~~~~~~~~~~~~~~~~~~~~~~~~~~~~~~~~~~~~~~~~~- \frac{e^2 \lambda_{V}^2}{m_\pi^2 m_V^2}
\frac{d}{dt} \left[ \dot\zeta(t)  \langle A_{{\rm T} }^i  {\vec \nabla}^2  A_{{\rm T} }^i \rangle(t) \right] =0
.
\label{eom}
\end{eqnarray}

Now we decompose the fields ${\vec \pi}$ 
and $\vec A_T$ into their Fourier
mode functions $U_{\vec k}(t)$ and $V_{\lambda \vec k}(t)$ respectively,
\begin{eqnarray}
{\vec \pi}(\vec x,t)&=&\int\frac{d^3k}{\sqrt{2(2\pi)^3\omega_{\pi\vec k}}}
\left[{\vec a}_{\vec k} U_{\vec k}(t) e^{i\vec k\cdot \vec x} + {\rm h.c.}\right],
\nonumber \\
\vec A_T(\vec x,t)&=&\sum_{\lambda=1,2} \int
\frac{d^3k\;\vec \epsilon_{\lambda \vec k}}{\sqrt{2(2\pi)^3\omega_{A\vec k}}}
\left[b_{\lambda \vec k} V_{\lambda \vec k}(t) e^{i\vec k\cdot \vec x} +
{\rm h.c.}\right],
\end{eqnarray}
where ${\vec a}_{\vec k}$ and $b_{\lambda \vec k}$ are destruction operators, and
$\vec \epsilon_{\lambda \vec k}$ are circular polarization unit vectors.
The frequencies $\omega_{\pi\vec k}$ and $\omega_{A\vec k}$ 
can be determined from the initial states and will be specified below.
Then the mode equations can be read off from the quadratic part 
of the Lagrangian in the form
\begin{eqnarray}
&&\left[\frac{d^2}{dt^2}+k^2 + m^{2}(t) +4 \lambda \phi^2(t) +4 \lambda \zeta^2 (t) + 4 \lambda\Sigma(t) \right]U_k(t)=0, \nonumber \\
&&\frac{d^2 }{dt^2} V_{1 k}(t) +\left[ 1- \frac{e^2 \lambda_V^2}{m^2_{\pi} m^2_V} \dot\zeta^2 (t) \right] k^2 V_{1 k}(t)-k \frac{e^2}{2\pi^2 f_{\pi} }  \dot{\zeta} 
(t) V_{1 k}(t)=0,  \nonumber \\
&&\frac{d^2 }{dt^2} V_{2 k}(t) +\left[ 1- \frac{e^2 \lambda_V^2}{m^2_{\pi} m^2_V} \dot\zeta^2 (t) \right] k^2 V_{2 k}(t)+ k \frac{e^2}{2\pi^2 f_{\pi} }  \dot{\zeta} 
(t) V_{2 k}(t)=0. 
\label{meq}
\end{eqnarray}
With $\lambda_V=0$, this reproduces the mode equations derived in 
Ref.~\cite{boy2}.
To solve these mode functions, we must specify initial conditions.
At the time of ``quench'', we assume that the quantum fluctuations for the $\pi$
fields which undergo  the strong interactions are in the local thermodynamic equilibrium at the initial temperature $T_i > T_c$ with the chiral order parameter displaced initially away from the equilibrium position and with a non-vanishing 
initial amplitude along the $\pi^0$ direction, i.e., $\zeta (0) \ne 0$ and $\phi (0) =0$.  However, since the photons interact 
electromagnetically,  their mean free paths are longer than the estimated size of the presumed quark-gluon plasma fireball  so that   the produced photons will escape from the plasma freely \cite{bell}.
Therefore, one can argue that    
 the photonic medium effects  play no role 
in the dynamics of photon production.  Based on the above arguments, the initial conditions for the mode functions are  given  by
\begin{eqnarray}
&&U_k(0)=1,\;\dot U_k(0)= -i\omega_{\pi k},
\;\omega_{\pi  k}^2 = k^2+m^2 (t< 0)+4 \lambda [\phi^2(0)+\zeta^2(0)]\;; \nonumber \\
&&V_{\lambda k}(0)=1,\;\dot V_{\lambda k}(0)= -i\omega_{A k},
\;\omega_{A k}=k,
\label{ic}
\end{eqnarray}
with the  expectation values with respect to the initial states  given by
\begin{eqnarray}
\Sigma(t)&=& (N-1) 
\int^\Lambda\frac{d^3k}{2(2\pi)^3\omega_{\pi k}} \left[ |U_k(t)|^2-1 \right] \coth \left[ \frac{\omega_{\pi k}}{2 T_i} \right]  , \nonumber \\
\epsilon^{\alpha\beta\mu\nu} \langle F_{\alpha\beta}   F_{\mu\nu} \rangle(t) 
&=&   \int^{\Lambda}  \frac{d^3k}{2(2\pi)^3\omega_{A k}} (4k) \frac{d}{dt} \left[ |V_{2 k}(t)|^2- |V_{1 k}(t)|^2 \right], \nonumber \\
 \langle A_{{\rm T} }^i  {\vec \nabla}^2  A_{{\rm T} }^i \rangle(t) &=&
 \int^\Lambda \frac{d^3k}{2(2\pi)^3\omega_{A k}} (-k^2)
\left[ |V_{1 k}(t)|^2+ |V_{2 k}(t)|^2 \right],
\label{qf}
\end{eqnarray}
where we set the cutoff scale $\Lambda\simeq m_V$ and  $N=3$.
The above specified initial conditions are physically plausible and  
simple enough for us to investigate a quantitative description of the dynamics.
The expectation
value of the number operator for the asymptotic photons with momentum 
$\vec k$ is given by \cite{boy2}
\begin{eqnarray}
\langle {\bf N}_{k}(t)\rangle&=&
 {\frac{1}{2k}} \left[ \dot{\vec A_T}( {\vec k},t) \cdot 
\dot{\vec A_T}( -{\vec k},t) \right. \left.
+k^2  {\vec A_T}( {\vec k},t) \cdot 
{\vec A_T}( {-\vec k},t) \right]-1 \nonumber \\  
&=& {1\over 4k^2}\sum_\lambda
\left[|\dot V_{\lambda k}(t)|^2+k^2
| V_{\lambda k}(t)|^2\right] - 1.
\end{eqnarray}
This gives the spectral number density of the photons produced at time $t$,
$dN(t)/d^3k$.

\begin{figure}
\leavevmode
\hbox{
\epsfxsize=6.5in
\epsffile{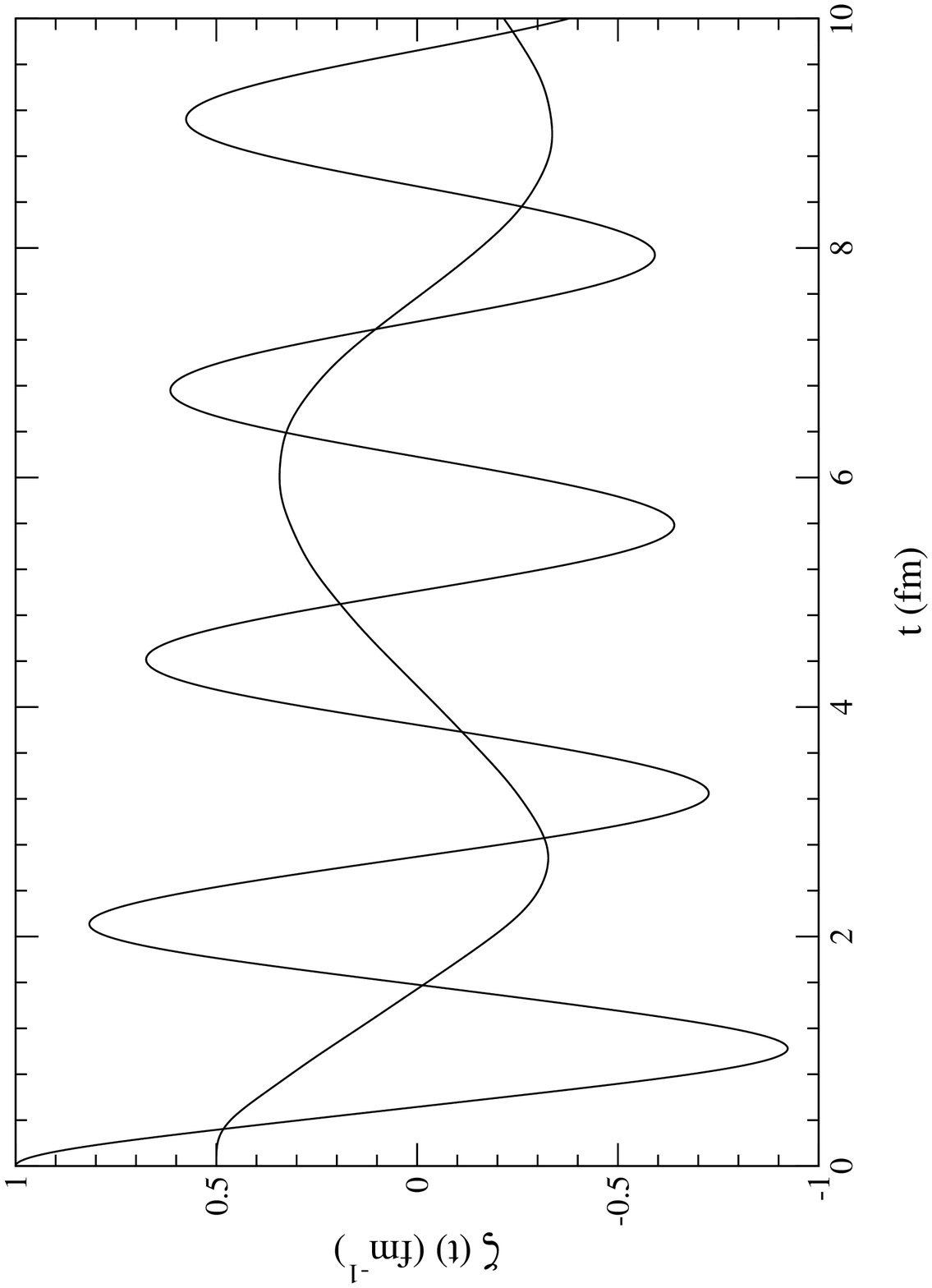}}
\caption{}
\label{fig1}
\end{figure}

\begin{figure}
\leavevmode
\hbox{
\epsfxsize=6.5in
\epsffile{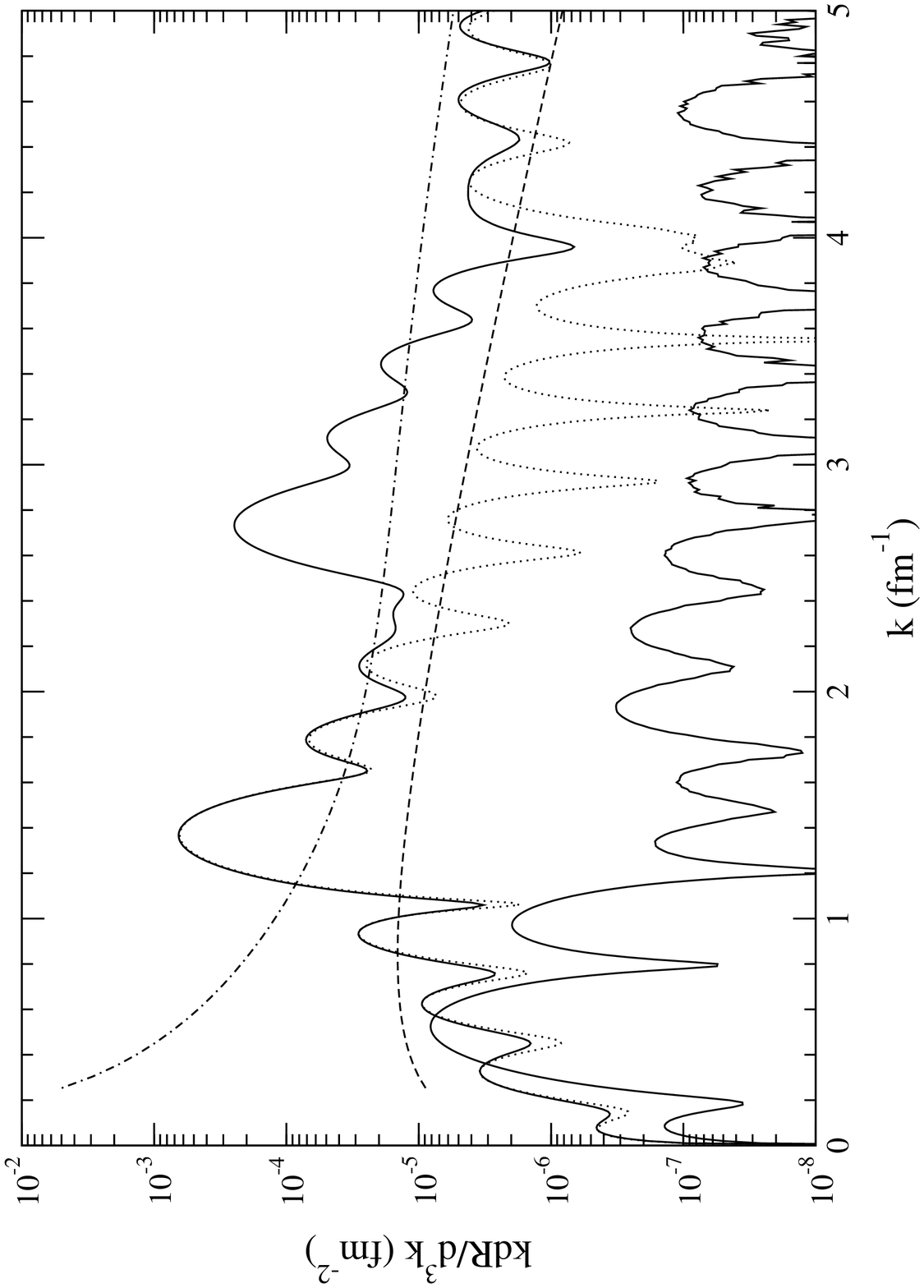}}
\caption{}
\label{fig2}
\end{figure}

We now perform the numerical analysis. 
We choose to represent the quench from an initial temperature    set to be $ T_i=220~{\rm MeV}=1.1 ~{\rm fm}^{-1}$ to zero temperature \cite{boy2}.  The evolution is tracked up to a time
of about $ 10 ~ {\rm fm}$ after which the hydrodynamical expansion becomes important.  
Fig.~\ref{fig1}  shows the temporal evolution of the $\zeta (t)$  by choosing the initial conditions $\zeta (0)=0.5~ {\rm fm}^{-1}$ and $1~ {\rm fm}^{-1}$, and $\dot{\zeta}(0)=\phi(0)=\dot{\phi}(0)=0$. The $\zeta (t) $ evolves with damping due to the
backreaction effects from the quantum fluctuations.  
In Fig.~\ref{fig2}, we present the time-averaged invariant photon production rate,
$kdR/d^3k$, where 
\begin{equation}
dR={1\over T}\int_0^T \frac{dN(t)}{dt} dt,
\label{dR} 
\end{equation}
over a period from the initial time to time $T=10~ {\rm fm}$.

In the case of $\zeta (0)=1~ {\rm fm}^{-1}$, the  
produced non-thermal photons
have spectrum peaks around two photon momenta, $k=1.35 ~ {\rm fm}^{-1}  $ 
and $k=2.7~  {\rm fm}^{-1} $, which exhibits the features of the unstable bands 
and the growth of the fluctuating modes.
The growth of the modes in the unstable bands translates into the profuse particle production. 
Note that   the peaks are located at $k=\omega_\zeta/2$ and $k=\omega_\zeta$ where $\omega_\zeta$ is the oscillating frequency  of the
$\zeta (t)$ field with $\zeta (0)=1~ {\rm fm}^{-1}$ in Fig.~\ref{fig1}.
The  spectrum peaks clearly result from the oscillations of the $\zeta$ field
that serves as the time dependent frequency term in  the mode equations of $\vec{A}_{T}$ (\ref{meq}).
Thus,  the   photon production mechanism is that of    parametric amplification.
Comparing  with the results in Ref.~\cite{boy2}, 
where the authors consider the photon production only via the $U_{\rm A} (1)$ anomalous interaction, 
and the dotted curve in Fig.~\ref{fig2} which denotes the 
time-averaged invariant photon production rate for $\zeta (0)=1~{\rm fm}^{-1}$ 
with the vector meson channel turned off ($\lambda_V=0$), 
we can easily recognize  
that  the $1.35~{\rm fm}^{-1}$ peak is resulted 
from  the coupling $\pi^0 F {\tilde F}$ while 
the $2.7 ~{\rm fm}^{-1}$ peak is from the interaction
$(\partial\pi^0 {\tilde F})^2 $. 
For a smaller initial field amplitude $\zeta (0)=0.5~ {\rm fm}^{-1}$, 
the oscillating frequency decreases while
the peaks shift to the lower-momentum
region with peak values almost two orders of magnitude lower than those of
$\zeta (0)=1~{\rm fm}^{-1}$. This explicitly shows the non-linearity of the
amplification process.

We now provide the analytical analysis to  understand qualitatively the above numerical results and
especially why the $(\partial\pi^0 {\tilde F})^2 $ coupling is 
dominant at $k= 2.7 ~{\rm fm}^{-1}$ in photon production although
it is small perturbatively. 
From Fig.~\ref{fig1} for $\zeta (0)=1~{\rm fm}^{-1}$, it shows that the solution of  $\zeta (t)$ is a quasiperiodic function with a decreasing  amplitude  during  the first few oscillations.
To obtain the analytical estimates for the locations of the unstable bands in the produced photon spectrum as well as their growth rates, let us approximate the   $\zeta (t)$ as
$\zeta(t)\simeq  \bar{\zeta} \sin(\omega_\zeta t)$.  
The $\bar{\zeta}$ is the average amplitude over a period from the initial time up to time of $10~{\rm fm}$, 
 and is about  $\bar{\zeta} \simeq 0.8 ~{\rm fm}^{-1}$, and the oscillation frequency, $ \omega_\zeta \simeq 2.7~{\rm fm}^{-1}$,  measured directly from Fig.~\ref{fig1}. 
Then, the photon mode equation
in Eq.~(\ref{meq}) becomes 
\begin{equation}
\frac{d^2 }{dt^2} V_{1 k}(t) +\left[ 1- 
\frac{e^2 \lambda_V^2 \bar{\zeta}^2 \omega_\zeta^2}{2 m^2_{\pi} m^2_V} 
\cos(2\omega_\zeta t) \right] k^2 V_{1 k}(t)-k 
\frac{e^2 \bar{\zeta} \omega_\zeta}{2\pi^2 f_{\pi} }  
\cos(\omega_\zeta t) V_{1 k}(t)=0.
\label{mathieu}
\end{equation}
When the vector meson channel is turned off ($\lambda_V=0$), we change the
variable to $z=\omega_\zeta t/2$. Then, Eq.~(\ref{mathieu}) becomes
\begin{equation}
\frac{d^2 }{dz^2} V_{1 k} + \frac{4k^2}{\omega_\zeta^2} V_{1 k} 
-\frac{4k e^2 \bar{\zeta}}{2\pi^2 f_{\pi} \omega_\zeta}
\cos(2z) V_{1 k} =0.
\end{equation}
This is the standard Mathieu equation~\cite{mac}. 
The widest and most important instability is the first parametric resonance 
that occurs at $k=\omega_\zeta/2$ with a narrow
bandwidth $\delta\simeq e^2 \bar{\zeta}/(2\pi^2 f_{\pi})$.
The instability leads to the exponential growth of photon modes with a growth
factor $f= e^{2\mu z}$, where the growth 
index $\mu\simeq \delta/2$. This growth explains the peak at 
$k=1.35 ~ {\rm fm}^{-1}$ in Fig.~\ref{fig2}.
When the ${\rm U}_{\rm A} (1)$ anomalous vertex is turned off 
($f_\pi \rightarrow \infty$), we change the
variable to $z'=\omega_\zeta t$. Then, Eq.~(\ref{mathieu}) becomes
\begin{equation}
\frac{d^2 }{dz'^2} V_{1 k} + \frac{k^2}{\omega_\zeta^2} V_{1 k} 
-\frac{ k^2 e^2 \lambda_V^2 \bar{\zeta}^2}{2 m^2_{\pi} m^2_V} 
\cos(2z') V_{1 k} =0.
\end{equation}
Now, the parametric resonance occurs at $k=\omega_\zeta$ with a growth factor
$f'= e^{2\mu' z'}$, where
$\mu'\simeq e^2 \lambda_V^2 \bar{\zeta}^2 \omega_\zeta^2/(8 m^2_{\pi} m^2_V)$.
This growth explains the peak at $k=2.7 ~ {\rm fm}^{-1}$ in Fig.~\ref{fig2}.
Taking $\bar{\zeta}=0.8~ {\rm fm}^{-1}$ and $t \simeq 10~{\rm fm}$,
the ratio of their growth rates is given by 
${\dot f'}/{\dot f}  \simeq 0.5$.
This means that the height of the $2.7~{\rm fm}^{-1}$ peak is about
one half of the $1.35~{\rm fm}^{-1}$ peak as we can see in Fig.~\ref{fig2}.

We then compare our results with the 
thermal photon emitted from a quark-gluon plasma and a 
hadron gas. 
In Fig.~\ref{fig2}, the invariant photon production rate for the quark-gluon
plasma is drawn using the parameters given in Ref.~\cite{kap}. 
For the hadron gas, we have used  the rates for
the most important scattering and decay processes~\cite{nadeau}.
It is shown that the $1.35~{\rm fm}^{-1}$ peak and
the $2.7~{\rm fm}^{-1}$ peak 
is about an order of magnitude larger than the thermal photons.
Therefore, we can come to the conclusion that these non-thermal photons can be regarded as a distinct signature 
of non-equilibrium DCCs. 

In conclusion, we have studied  the production of photons  through the non-equilibrium relaxation of a 
disoriented chiral condensate within which  the chiral order parameter  initially has a non-vanishing expectation value along the $\pi^{0}$ direction   . 
Under the ``quench'' approximation,
the invariant production rate for non-equilibrium photons driven by the 
oscillation of the $\pi^0$ field due to parametric amplification is given, 
which exceeds that for thermal photons from a thermal quark-gluon 
plasma or hadron gas for photon energies around $0.2-0.7\;{\rm GeV}$. 
These relatively high-energy non-thermal
photons can be a potential test of the formation of disoriented chiral condensates in   
relativistic heavy-ion-collision experiments.

\medskip
We would like to thank D. Boyanovsky for his useful discussions.
The work of D.S.L. (K.W.N.) was supported in part by the 
National Science Council, 
ROC under the Grant NSC89-2112-M-259-008-Y (NSC89-2112-M-001-001).

\newpage

\begin{center}
{\bf FIGURE CAPTIONS}
\end{center}
\bigskip
\noindent
Fig.~\ref{fig1}  
Time evolution of the mean field $\zeta(t)$ with initial amplitudes
$0.5~{\rm fm}^{-1}$ and $1~{\rm fm}^{-1}$.
\medskip

\noindent
Fig.~\ref{fig2} 
Lower and upper solid curves are the time-averaged invariant photon production 
rates for $\zeta (0)=0.5~{\rm fm}^{-1}$ and $1~{\rm fm}^{-1}$ respectively.
The latter shows profuse photon production.
The dot-dashed curve is the invariant photon production rate for the
thermal hadron gas at $T=220~{\rm MeV}$. The rate for $T=220~{\rm MeV}$
quark-gluon plasma is denoted by the dashed curve. Also shown is the dotted
curve which denotes the rate for $\zeta (0)=1~{\rm fm}^{-1}$ with the vector
meson channel turned off, i.e., $\lambda_V=0$. 
\medskip

\newpage

\newpage

\end{document}